\documentstyle[epsf]{mn}

\newcommand{\iras}{IRAS\ 04210+0400}
\newcommand{\kms}{{\rm\ km\ s}$^{-1}$}
\newcommand{\oiii}{[O\,{\sc iii}] 5007-\AA}

\title[Gas kinematics in \iras]{Kinematics of ionized gas associated with the
radio nucleus and lobes in the active galaxy \iras}
\author[A.J. Holloway et al.]
       {A.J. Holloway,$^{1}$ W. Steffen,$^{1}$ A. Pedlar,$^{2}$ D.J.
Axon,$^{2,3}$ J.E. Dyson,$^{1}$ J. Meaburn$^{1}$\cr and C.N. Tadhunter$^{4}$\\
    $^{1}$Department of Physics and Astronomy, University of Manchester,
Schuster Laboratory, Oxford Road, Manchester M13 9PL \\
 $^{2}$Nuffield Radio Astronomy Laboratories, University of Manchester, Jodrell
Bank, Macclesfield, Cheshire SK11 9DL \\
 $^{3}$ESA secondment, Space Telescope Science Institute, Baltimore, MD 21218,
USA\\
 $^{4}$Department of Physics, Hicks Building, University of Sheffield,
Sheffield S3 7RH}
\date{Accepted for publication in MNRAS.}

\begin{document}

\maketitle

\begin{abstract}

We have used high resolution longslit spectroscopy to investigate the
ionized gas in the active galaxy \iras\ and its association with the
radio structure.

We suggest that two of the ionized components are associated with the
central double radio source and observe that the relative positions of
these components vary for different emission lines. Both results are
consistent with the radio components representing the working surfaces
of a pair of jets emerging from the centre of the galaxy. In this
scenario, the optical emission in the centre arises behind the bowshocks
produced by the jets in the interstellar medium.

The emission lines are detected and show a dramatic ($\approx$ 900
\kms) spread in velocity at the position of the radio lobe
hotspots. We suggest a model which explains this phenomenon as the
result of a jet head emerging through the boundary between the
interstellar and intergalactic medium. A similar scenario has
previously been suggested as a model to explain wide angle tail radio
sources (WAT's).  Based on this model, we simulate the longslit
spectra of these regions and compare the results with the
observations.

\end{abstract}

\section{Introduction}

Ionized gas has been known to be associated with active galactic
nuclei (AGN), at least since Seyfert's discovery of broad lines from the
bright star-like nuclei in several nearby spirals.  There are at least two
distinct nuclear components to this gas consisting of the broad line region
(BLR) emitting permitted lines with velocity widths $\sim$3000
\kms, and the narrow (or forbidden) line region (NLR) with line widths
of $\sim$500 \kms.  The BLR gas appears to be confined to a region
less than a parsec in size at the centre of the galaxy (Rees, Netzer
\& Ferland 1989). The extent of the NLR is more difficult to quantify,
but there is no doubt that in typical Seyferts the strongest NLR
emission is confined to the central few hundred parsecs of the active
galaxy.  However, in the last decade high sensitivity observations of
active galaxies have revealed regions of ionized gas extending up to
around 10 kpc (e.g. Markarian 6, Meaburn et al. 1989).

There also are two distinct types of extended ionized regions. The
extended narrow line region (ENLR) is characterised by line widths
less than $100$ \kms\, and kinematics consistent with quiescent
rotating gas in the host galaxy. It has been proposed that it consists
of gas which which is dynamically undisturbed by the central activity,
although it is photoionized by the anisotropic UV emission from the
AGN (Unger et al. 1987). This model, which has been reinforced by the
detection of a number of `wedge' shaped ENLRs (e.g. NGC1068, Pogge
1988, Unger et al. 1992, NGC5252, Tadhunter \& Tsvetanov 1989, Mkn78,
Pedlar et al. 1989), is consistent with photo-ionization by a cone of
UV radiation from the nucleus.

The second type of extended ionized region (often loosely known as an
EELR - Extended Emission Line Region) also extends over approximately
10\,kpc but exhibits linewidths of several hundred \kms\ and shows
velocities largely unrelated to the dynamics of the host galaxy. Often
these regions are associated with radio jets (e.g. 3C120, Axon et
al. 1989; 3C305, Heckman et al. 1982, Jackson et al. 1995) and are
thought to represent the interaction of collimated ejection from the
nucleus with the interstellar medium (ISM) or the intergalactic medium
(IGM).  In many cases, the gas appears to be photoionized by UV
radiation from the nucleus, although it cannot be excluded that shock
ionization is also important. The study of these EELR provides
information on the parameters of host galaxy and its ISM and on those
of the collimated ejection from the AGN. This information can be
combined with the results from radio observations to further constrain
model calculations. In this paper we report on a study of ionized gas
associated with this latter type of extended emission line region.

The galaxy \iras\ was first detected in scans of the Infrared
Astronomical Observatory (IRAS) at 25 and 60 microns (Soifer et
al. 1984). It has a redshift of z=0.0462 (Beichman et al. 1985)
implying a distance of 185\,Mpc, at which 1\arcsec\, corresponds to
900\,pc (assuming {\it H}$_{0}$=75 \kms\,Mpc).  Further work by
Beichman et al. (1985) identified the IRAS source with a spiral galaxy
having an integrated R band magnitude of 16.3.  Their spectroscopy
revealed a Seyfert type-2 emission-line nucleus, although its
association with extended radio emission was unusual. Hill et
al. (1988) showed the radio structure to consist of a central double
source and large scale ($\sim$ 25\,kpc) double radio lobes extending
beyond the optical galaxy.  Typically the nuclei of spiral galaxies,
including Seyferts, are radio quiet objects, with radio luminosities
of $\sim 10^{20-21}$ Watts~Hz$^{-1}$ at 20\,{\rm cm} wavelength.  If
radio emission is observed, the structures are usually smaller than
600\,pc (e.g. Ulvestad \& Wilson 1984). Although radio `lobes' are not
unknown in spiral galaxies (e.g. NGC~3079 and NGC~5548), \iras\ would
be a rare example of a spiral galaxy with associated radio structure
more like that of an FR1 radio galaxy with hotspots and diffuse
lobes. However Hill et al. (1988) have questioned the classification of
\iras\ as a spiral and suggested that the `spiral arms' may be
associated with the radio ejecta. The observed 20\,cm radio luminosity
of $2.4 \times 10^{23}$ Watts Hz$^{-1}$, the large radio lobes, and
the narrow emission line spectrum more nearly fits the definition of a
Narrow Line Radio Galaxy (NLRG) which are associated with elliptical
galaxies.

The radio galaxies 3C\,305 (Heckman et al. 1982, Jackson et al. 1995),
4C\,29.30 (van Breugel et al. 1986), and 4C\,26.42 (van Breugel, Heckman \&
Miley 1984) have similar radio and optical properties. They all show
radio jets which, starting from hotspots, flare and bend into extended
radio lobes. Nevertheless, \iras\ is distinguished from these cases
through the alignment of the bent lobes with the spiral structure and
the high symmetry of its radio and optical emission.

In this paper we concentrate on the kinematics and spatial positions
of the spectral features.  In Section 2 we present the imaging and
spectral observations obtained at the William Herschel Telescope (WHT)
and at the Isaac Newton Telescope (INT) on La Palma, respectively. In
Section 3 we discuss the observational results and outline a model
which reproduces the main spectral features. Our conclusions are
summarised in Section 4.

\section{Observations and results}

\subsection{WHT Imaging}

An 800 second exposure R-band image of the galaxy was obtained using
the 4.1m William Herschel Telescope on 1993 December 13. An EEV
CCD was used at the auxiliary focus with the 22 micron square pixels
providing an image scale of 0.10 arcsec/pixel. The image was processed
in the usual manner but not flat fielded. This is not a serious
problem however as the area of interest is small and the image was not
to be used for photometry. The seeing during the observation was
typically 1.6 arcsec and the images were also blurred by a slight
shift in focus of the telescope during the exposure.

The image is shown in Fig. \ref{wht} with the radio map contours and
slit position overlayed. It is displayed on a logarithmic
intensity scale to enable the bright core and faint spiral features to be
seen.

\begin{figure*}
\vspace{9in}
\caption{WHT R-band image displayed with logarithmic scaling and VLA 6\,cm
radio contours (Hill et al. 1988) superimposed.}
\label{wht}
\end{figure*}

\subsection{INT Spectroscopy}

Long-slit spectra were obtained using the Intermediate Dispersion
Spectrograph (IDS) and GEC7 CCD at the Cassegrain focus of the 2.5m
Isaac Newton Telescope on La Palma over the nights of 1993 December 4
and 6.  The 500mm camera was used to provide a spatial scale of 0.3
arcsec pixel$^{-1}$ and a wavelength dispersion of 0.48\AA\
pixel$^{-1}$ over the 590$\times$400 pixel array.  The two exposures
were taken at position angle 0$^{\circ}$ centred on the nucleus to
coincide with the radio axis. Observing conditions varied for the
exposures; the \oiii\ region centred on 5160\AA\ was observed for 3600
seconds under 1.6 arcsec seeing and the H$\alpha$ region centred on
6950\AA\ was observed for 6000 seconds with 1.0 arcsec seeing and a
0.7 arcsec slit width. The wavelength resolution as determined from
the arc spectra, is 1.1\AA\ (63\kms\ at [O\,{\sc iii}] 5007\AA\ and
48\kms\ at H$\alpha$).

The data were processed at the University of Manchester node of the UK
STARLINK network using programs from the {\sc figaro, twodspec}, and
{\sc kappa} packages.  The bias level of the CCD was measured from the
overscan region of the chip and a mean value subtracted.  The spectra
were calibrated in wavelength and corrected for curvature along the
slit length using the copper-neon arc lamp spectrum. Once wavelength
calibrated, the background sky lines were removed by averaging two
strips which were clear of object data on either side of the continuum
and subtracting this sky spectrum from each row. The spectra were
corrected for the instrumental response and flux calibrated using the
standard star HD\,93521.

\begin{figure*}
\centering
\mbox{\epsfxsize=6in\epsfbox[8 18 566 369]{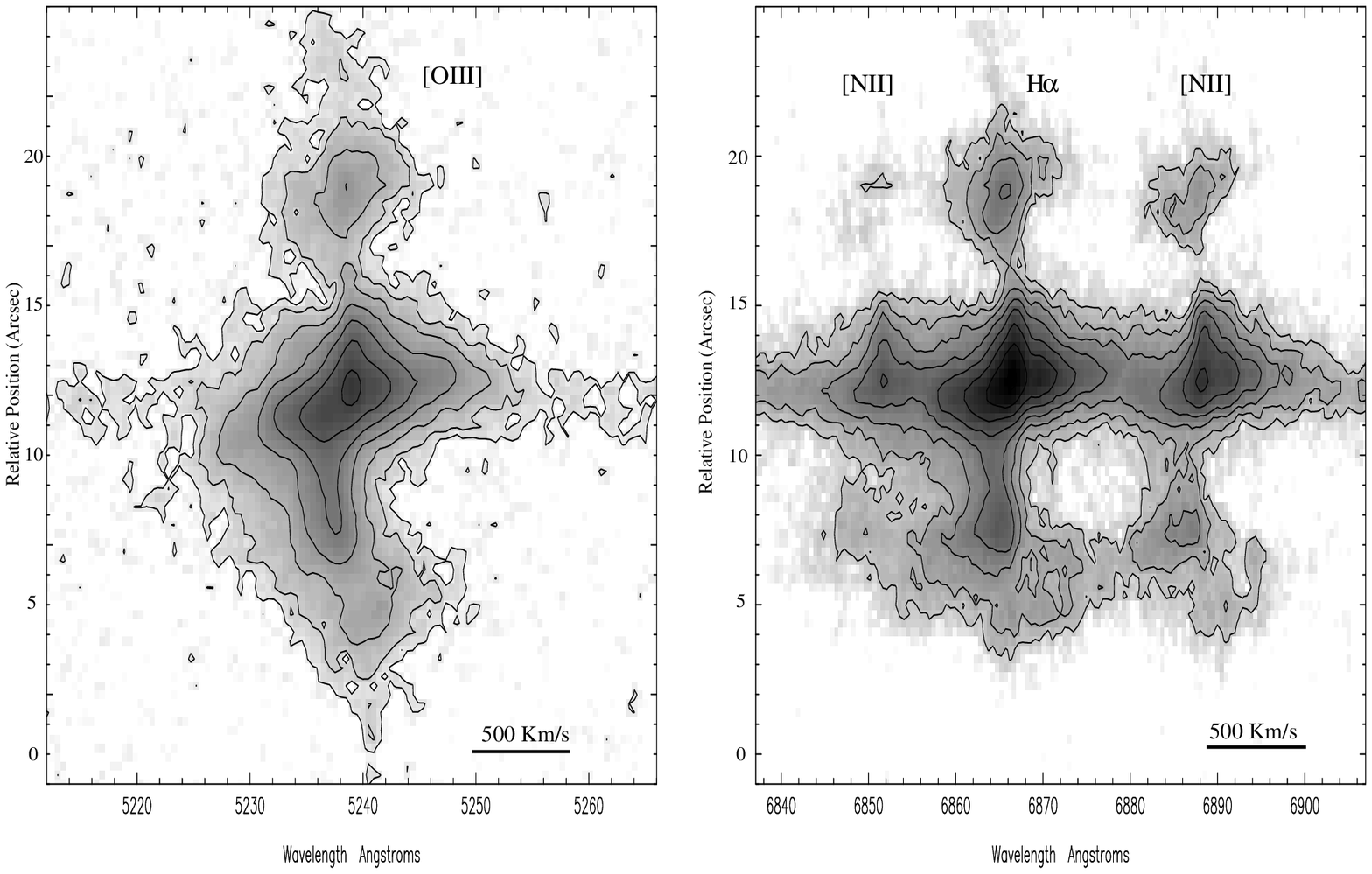}}
\caption{INT IDS longslit spectra, displayed on a logarithmic scaling, for the
wavelength regions containing (a) the \oiii\ line with contours at 1, 2, 5, 10,
20, 40 and 80 per cent of the peak intensity and (b) H$\alpha$ + [N\,{\sc ii}]
6548,6584-\AA\ lines with contours at 1, 2, 5, 10, 20, 40, 70 and 90 per cent
of the peak intensity.}
\label{int}
\end{figure*}

\begin{figure*}
\centering
\mbox{\epsfclipon\epsfxsize=6in\epsfbox[0 272 540 509]{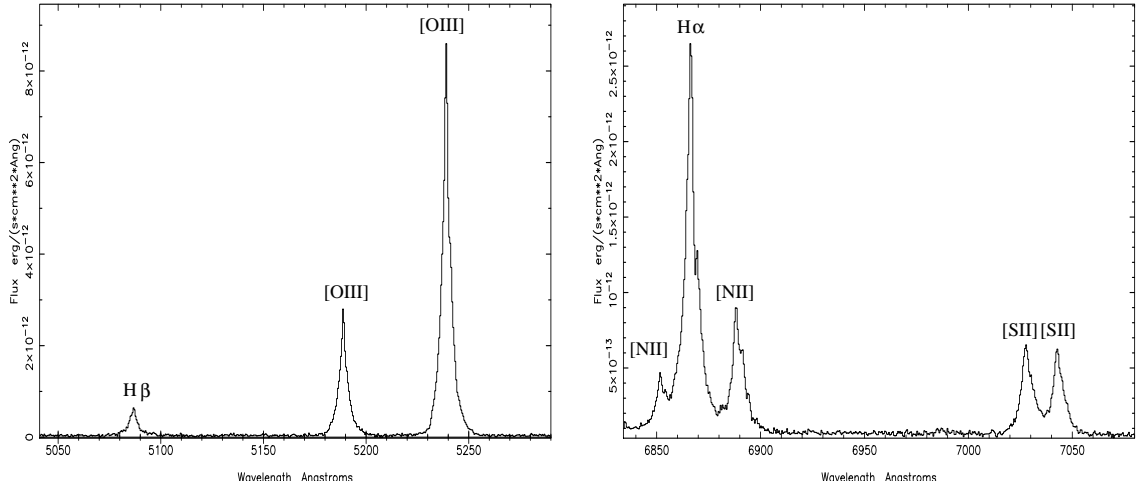}}
\caption{INT IDS spectra of the nucleus of \iras\ over the wavelength ranges
(a) 5040-5290\AA\ and (b) 6835-7080 \AA.}
\label{nucspec}
\end{figure*}

Position-velocity maps of the \oiii\ and H$\alpha$+[N\,{\sc ii}] lines
are presented in Figs. \ref{int}a and \ref{int}b. They are displayed
with a logarithmic intensity scaling to enable both bright and faint
features to be seen clearly.  The \oiii\ line is broad ($\sim$400\kms)
close to the nucleus and the more extended gas has linewidths of
$\sim$800\kms. The H$\alpha$+[N\,{\sc ii}]6548,6584-\AA\ plot shows
similar velocities to the \oiii, but of particular note are the large
linewidths (900 \kms) seen in H$\alpha$ approximately 7\arcsec\, to
the south of the nucleus. In the nucleus we can detect a number of
fainter lines in spectra produced by co-adding data over the central
1\arcsec . H$\beta$, [O\,{\sc iii}] 4959-\AA\ and \oiii\ lines are
visible in Fig. \ref{nucspec}a and [N\,{\sc ii}] 6548-\AA, H$\alpha$,
[N\,{\sc ii}] 6584-\AA\ and [S\,{\sc ii}] 6716,6731-\AA\ lines are visible in
Fig. \ref{nucspec}b.

\section{Discussion}

In this section we discuss the observations, focusing on the
classification of the galaxy from broad band imaging (Section 3.1),
the peculiar gas kinematics observed in the core (Section 3.2) and in
the regions around 5\arcsec\ north and south of the core (Section
3.3). In Section 3.4 we outline a simple model which reproduces the
peculiar extended features found in the longslit emission line
profiles.

\subsection{The classification of \iras}

In our R-band image (Fig. \ref{wht}), we have detected emission from
\iras\ out to a distance of 8.5\arcsec\, and 9.5\arcsec, north and
south of the nucleus respectively. This gives the projected size of
the galaxy to be 16 kpc in diameter.  The central bulge-like region
has a radius of $\sim$ 2\arcsec (1.8 kpc) and the spiral features
extend a further $\sim$ 7\arcsec (6.3 kpc). To the north east of \iras\
lies a possible companion galaxy, this association being suggested by
its redshift of z=0.047 (Hill et al. 1988) which is similar to that of
\iras. The most luminous part of this companion is located 11\arcsec
(10 kpc) from the centre of the galaxy and the faint extended emission
extends over 7\arcsec (6.3 kpc) diametrically away from \iras. At
6.5\arcsec (5.9 kpc) west of \iras\ there is a faint object which may
also be interacting with the galaxy and further spectroscopy will be
required to check that it is not simply an object along our line of
sight.

The spiral features originate from the central region at position
angle $\sim$30$^{\circ}$ and appear to terminate at position angle
$\sim$0$^{\circ}$, in the vicinity of the 6cm VLA radio lobe hotspots
north and south of the nucleus. The radio lobes appear to continue the
curve defined by the arms out to a distance of 17\arcsec(15.3 kpc) and
14\arcsec(12.6 kpc), north and south respectively. If the curve
defined by the arms is continued to the limit of the radio emission,
the position angle of this point is $\sim$-30$^{\circ}$.

If the galaxy is a spiral, then the position of the spiral arms with
respect to the radio lobes is a coincidence and the extended rotation
pattern of the lobes is then explainable as a product of the galactic
rotation, assuming the jet to be near the disc of the galaxy. Pressure
bending as the jet leaves the galaxy environment could also lead to
the bending of the radio lobe, though then the observation that both
optical and radio features follow a similar curve would be a
coincidence.

If, on the other hand, the galaxy is an elliptical, then a different
mechanism to produce the spiral features is required. Such a model was
suggested by Hill et al. (1988) in which the spiral features are the
photoionized remnants of the radio jet. The clear alignment between
the nuclear double and the hotspots of the extended lobes
would suggest that the radio jet moves along this line and not in a
curved path along the spiral features.  Material swept up by the
passage of the jet would then be photoionized by the central UV source
and has since moved by rotational motions. This explains the alignment
of the spiral photoionized jet remnant and the radio lobes and
provides an argument against the idea of the extended features being
tidal tails.

The exact nature of the spiral features could be decided by further
spectroscopy of the region. As a working hypothesis, in the following
discussion we take the observed galaxy to be an elliptical, with
photoionized jet remnants.

\subsection{The nucleus}

The integrated spectrum of the nuclear region (Fig 3) shows FWHM
linewidths for the \oiii, H$\alpha$ and [S\,{\sc ii}] lines of 400\kms,
300\kms\ and 330\kms\ respectively. These are typical for the NLR of
Seyfert 2 and Narrow Line Radio galaxies.  We see no evidence of
broader permitted lines (representative of a broad line region, BLR).

In our longslit spectra, asymmetric spatial structure is found in the
core region (Fig. \ref{int}).  We identify several separate components
in velocity and space. Their positions, relative to the peak of the
continuum, are shown in Table 1.

The brightest peak of \oiii\ emission lies 0.51\arcsec\ to the north
of the continuum. The broad blue and red shifted components are
located 0.47\arcsec\ south and 0.78\arcsec\ north of the continuum
peak, respectively. The intensity peak of H$\alpha$, [N\,{\sc ii}], and
[S\,{\sc ii}] are found to be closer to the galaxy core than those of
the corresponding \oiii\ components. Though the seeing for the \oiii\
observation (1.6\arcsec) was considerably poorer than that for the
others (1\arcsec), we exclude this as a possible explanation of the
difference as the components are well separated in velocity space and
do not contaminate each other.

\begin{table}
{\centering
\caption{Separations of central components in different spectral lines,
         relative to the continuum position ($\pm 0.02$\arcsec). For
         comparison, the separation between the central radio
         components is 0.71\arcsec. Positive separations are to the
         north.}
\label{seps}
\begin{tabular}{lccc}
 Region    & Red comp.          &     Peak comp.      & Blue comp.\\
{[O\,{\sc iii}] 5007\AA}      & 0.78$\pm$0.05\arcsec & 0.51$\pm$0.05\arcsec &
-0.47$\pm$0.05\arcsec\\
H$\alpha$ & 0.41$\pm$0.03\arcsec & 0.37$\pm$0.03\arcsec &
-0.27$\pm$0.03\arcsec\\
{[N\,{\sc ii}] 6584\AA}   & 0.50$\pm$0.03\arcsec & 0.29$\pm$0.05\arcsec &
-0.11$\pm$0.05\arcsec\\
{[S\,{\sc ii}] 6716\AA}   & 0.45$\pm$0.05\arcsec & 0.34$\pm$0.06\arcsec &
-0.14$\pm$0.03\arcsec\\
\end{tabular}
}
\end{table}

We find systematic shifts between peak positions of the
different lines. The order of appearance of peaks is the same for the
northern and the southern components except for [N\,{\sc ii}]. Separations
to the south are smaller than to the north.

\begin{figure}
\centering
\mbox{\epsfxsize=3.25in\epsfbox{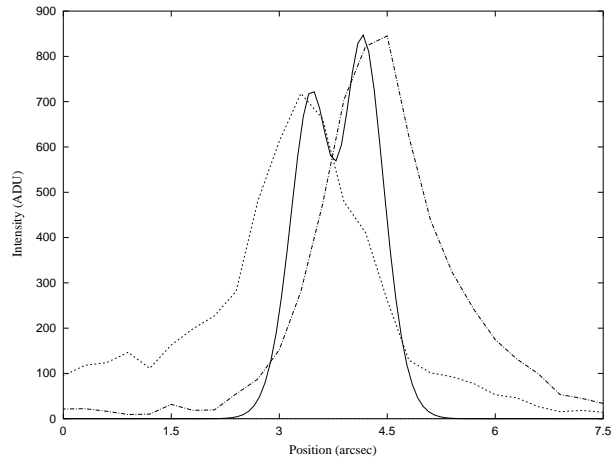}}
\caption{Plot of a two-component Gaussian fit to the
central radio intensity profile (Hill et al. 1988, solid line) and
cuts through blue (dotted) and red (dashed) components from the \oiii\
emission.  The wavelength range of the cuts is 5232-5236\AA~ for the
blue and 5245-5248\AA~ for the red component (see Fig. 2a). }
\label{nucsep1}
\end{figure}

\begin{figure}
\centering
\mbox{\epsfxsize=3.25in\epsfbox{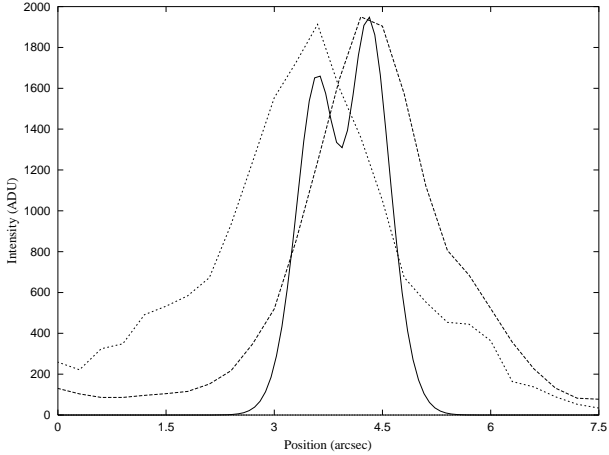}}
\caption{Plot of a two-component Gaussian fit to the
central radio intensity profile (Hill et al. 1988, solid line) and
cuts through normalised continuum (dotted) and the peak intensity (dashed)
components from the \oiii\ emission. The wavelength range of the cuts
is 5105-5167\AA~ for the continuum and 5238-5240\AA~ for the peak
component (see Fig. 2b).}
\label{nucsep2}
\end{figure}

In Figures \ref{nucsep1} and \ref{nucsep2} we show cuts through the
core sections of the \oiii\ longslit spectra in the spatial direction,
showing the intensity along position angle 0$^{\circ}$. In these plots
the relative strengths of the cuts is arbritary and only the spatial
information is used.  We include a two-component Gaussian fit to radio
data obtained from a 6cm VLA map which shows a compact double source,
oriented along postion angle 0$^{\circ}$ with a separation of
0.71\arcsec(resolution 0.55\arcsec, Hill et al, 1988). The position of
the radio with respect to the optical data is not exactly known so we
have adjusted the position to be symmetric with respect to the broad
blue and red shifted components, following our suggestion that they
are related to each other.

The association of radio and \oiii\ emission in the NLR of Seyferts
has been established in a number of cases (e.g. NGC5929 Whittle et
al. 1986, Mkn78 Pedlar et al. 1989). Models have been developed to
account for the association in which the radio plasma interacts with
the ISM in the bowshocks of expanding plasmons or working surfaces of
jets (Pedlar, Unger \& Dyson 1985, Taylor, Dyson \& Axon 1992, and
Wilson \& Ulvestad 1987). Based on our observations, we suggest that
this scenario also applies to the central region of \iras.

However, the separation between the location of the \oiii\ and
H$\alpha$ regions when comparing the red and blue components is
$\sim$0.3\arcsec (300 pc).  If these components are produced by the
suggested bowshocks, photoionization is not able to explain the large
difference in position of these spectral lines, taking into account
the high particle density of $\sim$10$^{4}$cm$^{-3}$ (from the [S\,{\sc
ii}] 6716,6731-\AA\ ratio) and the inferred bowshock velocity of
$\sim$300\kms.  Instead we suggest that the gas is mainly
collisionally ionized, which is consistent with the strength of the
[S\,{\sc ii}] 6716,6731-\AA\ lines. In this view the difference in
position of the \oiii\ and H$\alpha$ regions arises from the decrease
in shock speed normal to the bowshock surface at larger distances from
the apex. According to Cox \& Raymond (1985) the \oiii\ emission
dominates over the H$\alpha$ emission at shock speeds higher than
$\sim$100\kms, while the reverse is true for lower shock speeds. In
our case the H$\alpha$ emission would come from positions on the
bowshock where the normal to the surface makes angles of
$\geq$70$^{\circ}$ with the direction of propagation of the apex.

This interpretation does not account for the peak spectral component
located at 0.5\arcsec (for \oiii) north of the continuum
(Fig. 5). Within the error, this separation is identical to the
separation between the radio components. Therefore, we cannot exclude
the possibility that this strong feature is associated with the
northern radio component.  In this case the southern radio component
could represent the core of the galaxy and be associated with the
continuum emission.

At present there is no spectral index information on the nuclear
double and hence one of the radio components could be a compact flat
spectrum core which would be associated with the optical continuum
nucleus. Radio observations at two different frequencies will be
carried out to determine this.  However, the high overall structural
and spectral symmetry in \iras\ (analogous to NGC 5929 and other
Seyferts, in which flat spectrum cores are rare) leads us to adopt the
working hypothesis that the two radio components are associated to the
heads of a two sided jet.

\subsection{Extended Emission Line Regions}

An important result of our observations is the spatial and spectral
resolution of ionized gas with high velocity extending north and south
more than 10 kpc from the nucleus (Figs. 1 \& 2). The basic structure
can be appreciated best in the [N\,{\sc ii}] 6584-\AA\ line
(Fig. 2b). The northern hotspot at around 5\arcsec\ separation from
the centre shows a clearly ``V''-shaped structure, where the
redshifted arm is stronger and more extended.  In H$\alpha$ and \oiii\
a corresponding structure is found, although the blueshifted arm is
less obvious.

The southern hotspot (at approximately the same core separation) shows
a very similar, but inverted, structure, though even better defined:
most of the emission spreads into a blue wing, with increasing
negative relative velocity with a larger separation from the core. At
a lower brightness level, emission is seen spreading into a red
wing. Although the structure is marginally resolved, the observation
is suggestive of two components of acceleration or possibly a
ring-like structure in the velocity-space map. We find a considerable
difference in the \oiii\ line (Fig. 2a) in that the inverted
``V''-shape to the south is not obvious and instead all or most
emission seems to be in the redshifted wing.

There are three simple possible kinematic interpretations of such
``V''-shaped structures in the velocity map. They may arise from real
acceleration of the gas. Alternatively the gas might have a certain high
velocity when the line emission starts and the direction
of the velocity vector just changes. Naturally, the third possibility is a
combination of both effects.

\subsection{Modelling the spectra of the hotspots}

Jets in radio galaxies of moderate luminosities ($< 10^{25}$ \, {\rm
W\,Hz}$^{-1}$) can flare in only a few jet diameters and show very
large opening angles up to $90^{\circ}$ into diffuse lobes or tails
(O'Donoghue et al. 1993).  These structures often bend very near the
transition point as is the case for \iras\ (see
Fig. \ref{wht}). Norman et al. (1988) and Loken et al. (1995) have
modelled this phenomenon to explain the structure of wide angle tail
radio galaxies (WAT) in terms of an initially moderately supersonic
jet (Mach number $2-5$) jet passing through a shock or contact
discontinuity in the ambient gas where the jet flow becomes
subsonic. The jet is then disrupted and entrains external gas, which
becomes turbulent and large and small scale eddies then develop. Such a
shock in the ambient medium could be due to a supersonic galactic wind
moving into the surrounding intergalactic medium.

\begin{figure*}
\centering
\mbox{\epsfxsize=6in\epsfbox[0 93 498 403]{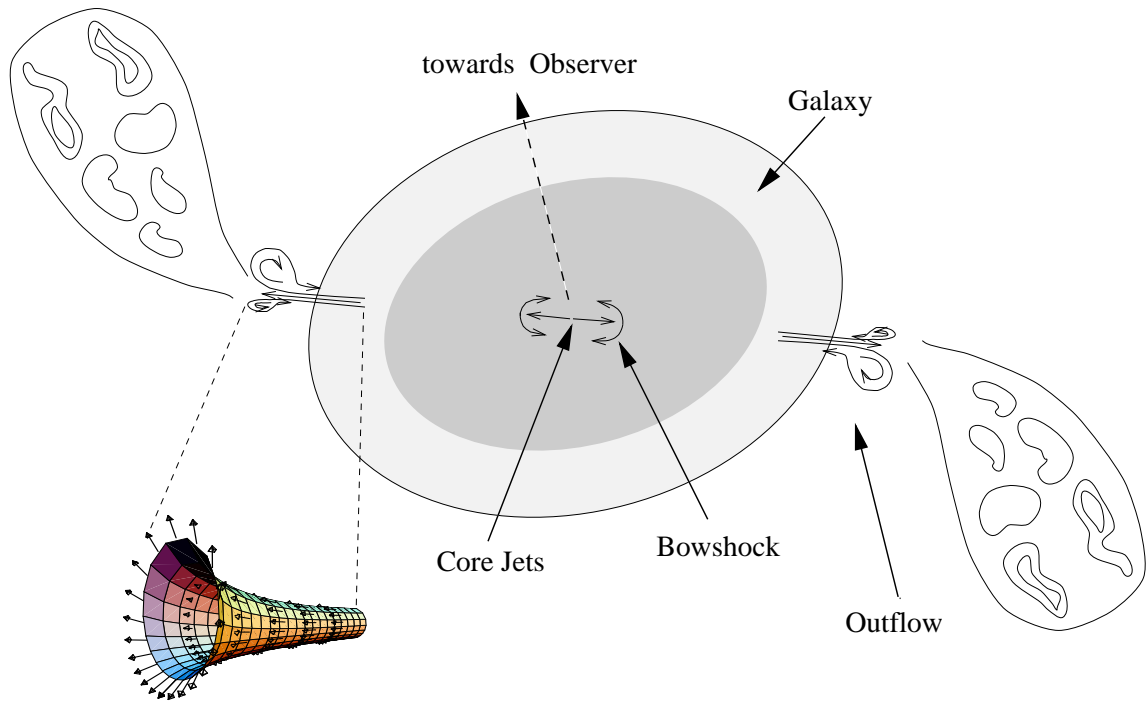}}
\caption{A schematic view of a model for \iras\ with a
central double bowshock and an expanding outflow for
the emergent jets at the galaxy/IGM boundary.}
\label{model}
\end{figure*}

We suggest that a similar scenario applies to \iras\ at the position
of the radio hot spots. A supersonic jet propagates through
the interstellar medium and passes through a transition region at the
edge of the galaxy (see Fig. \ref{model}). It emerges from the main
galaxy at the position of the hotspots. Here it goes subsonic,
disrupts and expands with a large opening angle producing the bent
conical lobes.  Ambient gas is entrained and accelerated
to several hundred kilometres
per second, thereby radiating in the observed emission lines.  In the
case of \iras\ this region is about 2.5\arcsec (2.3 kpc) large
starting at 4.5\arcsec (4 kpc) separation from the core. This is the
extension of both the enhanced optical and strongest radio
emission.

It is worth noting that Owen et al. (1990) conducted a search for
optical line emission from these flaring regions in WAT sources but
found no significant emission from the 5 objects they studied.

We model the longslit emission line spectra using a simple
parameterised description of the emission and velocity field of the
ionized gas flow. Fig. \ref{model} shows a schematic view of the
model.  Here we will give an outline of our model. A full account of
the theoretical model will be given in a forthcoming paper
(Steffen et al. in prep).

We concentrate on the kinematics of the hotspot positions and model
the emission line source as a collimated outflow which flares when
passing through the boundary between the ISM and the IGM (see
Fig. \ref{model}). The outflow opens gradually to an effective half
opening angle of $\sim$45$^{\circ}$ (where the exponentially decaying
emissivity becomes negligible). The longslit line profile is most
dependent on the orientation of the outflow with respect to the
observer's line of sight and the orientation of the spectrometer
slit. The detailed shape of the outflow and the change in emissivity
as a function of distance from the starting point only influence the
detail of the simulated spectrum and not the gross features.

The velocity is assumed to be constant (600\kms in the southern and
350\kms in the northern outflow) such that the structure found in the
longslit spectra results from the change in flow direction of the
gas. For simplicity, we assume the flow to be concentrated in a sheet
of Gaussian transverse emissivity distribution as shown schematically
in Fig. \ref{model}.

\begin{figure*}
\centering
\mbox{\epsfxsize=6in\epsfbox[51 159 382 474]{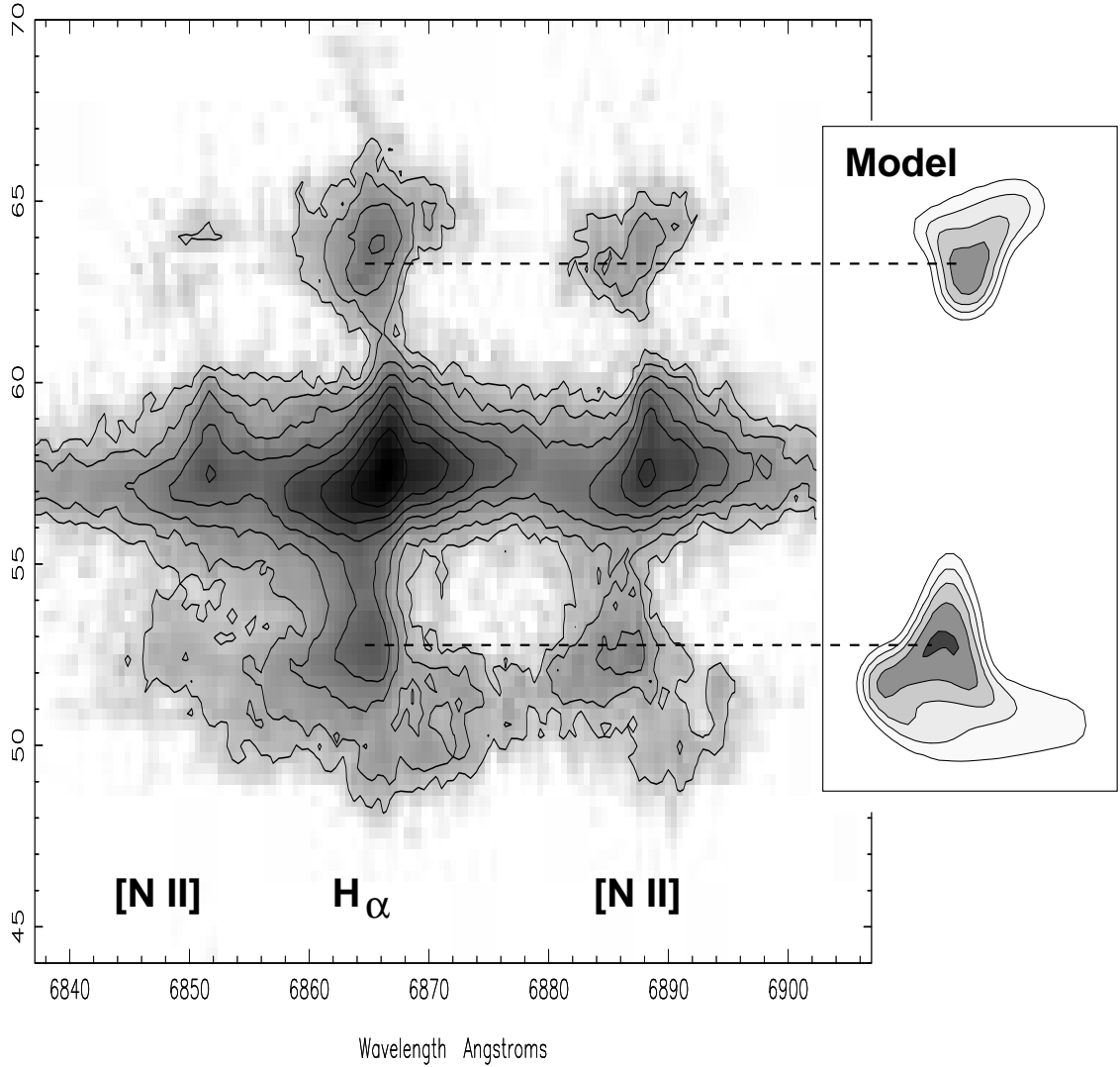}}
\caption{A comparison between the observed spectra and that generated by the
outflow model for the emerging jets.}
\label{modspec}
\end{figure*}

In Fig. \ref{modspec} we compare the observed H$\alpha$ and [N\,{\sc ii}] line
complex with our model. The axes of the outflows are inclined to the
sky plane at an angle of 15$^\circ$ (north) and 20$^\circ$ (south),
both towards the observer such that the direction of propagation is
not exactly colinear. This is suggested by the slight blueshift of
both hotspots with respect to the central region.  The slit direction
is north-south and the axes of the outflows are oriented at position
angles -25 $^\circ$ (north) and 150$^\circ$ (south) with respect to
the slit direction (north-south) consistent with the observed directions
of the radio lobes.  The slit is suficiently broad to
collect all the emission from the outflows.

The simulations reproduce the basic features of the observed spectral
line structure of the hotspot regions.  Thus, the optical observation
and modelling the interaction of jets with the transition region
between the ISM and the IGM could be an important tool to study
several aspects of the galaxy environment. In particular it provides
information about the velocities in the extragalactic jets which are
subject to debate. Otherwise, such direct kinematic information is not
available on the kiloparsec scale.

\section{Conclusions}

We have measured the spatial and velocity structure of the emission features
in the nucleus and coincident with the radio hotspots in the lobes.

\begin{enumerate}
\item We have observed a Seyfert type 2 spectrum for the nucleus
of \iras\ and with the evidence for it being an elliptical galaxy, it
would suggest that we are looking at a Narrow Line Radio Galaxy,
albeit with apparent spiral photoionized jet remnants.

\item In the nucleus we can identify broad velocity components with a similar
spatial separation as the double radio source. This suggests that we
are looking at a two sided jet with emission associated with the radio
jets. The observations can be explained by a bowshock model with
collisional excitation.

\item We see extended emission features with high velocity dispersion at
the position of the radio hot spots in the lobes. This is consistent
with a jet undergoing expansion in the IGM. We have modelled this
expansion as a jet which passes through the ISM/IGM boundary. Our
expanding outflow model reproduces the observed longslit spectra in
these regions.

\item This object provides a unique opportunity to study the interaction of
a jet with an ISM/IGM boundary at both optical and radio
wavelengths. This gives kinematic and morphological information which
so far has not been obtained for WAT sources which show a similar
phenomenology in the radio (though on a larger scale than here).

\end{enumerate}
\section*{Acknowledgements}

AJH and WS acknowledge the receipt of a PPARC studentship and PPARC
research associateship respectively. Thanks to the support staff of
the INT and WHT on these observing runs. We thank R.J.R. Williams and
A. Wilkinson for useful discussions.

\end{document}